\documentclass{jkas}

\def\year{2016} 
\def\month{September} 
\def\volume{00} 
\def\issue{0} 
\def\beginpage{1} 
\def\endpage{99} 
\def\received{---} 
\def\accepted{---} 

\date{Received \received ; accepted \accepted}

\def\muas{{\mu\rm as}}

\def\mas{{\rm mas}}
\def\lim{{\rm lim}}

\def\apj{{ApJ}}
\def\aj{{AJ}}

\def\jkas{{JKAS}}

\title{
     TGAS Error Renormalization from the RR Lyrae Period-Luminosity Relation
}

\author[1,2,3]{Andrew Gould}
\author[4]{Juna A. Kollmeier}
\author[1]{Branimir Sesar}

\affil[1]{Max-Planck-Institute for Astronomy, K\"onigstuhl 17, 69117 Heidelberg, Germany; \email{gould@astronomy.ohio-state.edu }}
\affil[2]{Korea Astronomy and Space Science Institute, Daejon 305-348, Republic of Korea}
\affil[3]{Department of Astronomy Ohio State University, 140 W.\ 18th Ave., Columbus, OH 43210, USA}
\affil[4]{Observatories of the Carnegie Institution of Washington, 813 Santa Barbara Street, Pasadena, CA 91101}

\def\corrauthor{
Andrew Gould
}

\def\runningauthor{
Gould \& Kollmeier

}

\def\runningtitle{
    TGAS Error Renormalization
}

\def\keywords{
astrometry
}

\def\abstracttext{
The Gaia team has applied a renormalization to their
internally-derived parallax errors $\sigma_{\rm int}(\pi)$
$$
\sigma_{\rm tgas}(\pi) = \sqrt{[A\sigma_{\rm int}(\pi)]^2 + \sigma_0^2}
\qquad (A,\sigma_0) = (1.4,0.20\,\mas)
$$
based on comparison to Hipparcos astrometry.
We use a completely independent method based on the RR Lyrae
$K$-band period-luminosity relation to derive a substantially
different result, with smaller ultimate errors
$$
(A,\sigma_0) = (1.1,0.12\,\mas) \qquad (\rm this\ paper).
$$
We argue that our estimate is likely to be more accurate and
therefore that the reported TGAS parallax errors should be reduced
according to the prescription:
$$
\sigma_{\rm true}(\pi) = \sqrt{(0.79\sigma_{\rm tgas}(\pi))^2 - (0.10\,\mas)^2}.
$$
}

\begin{document}
\jkashead 


\section{Introduction \label{sec:intro}}

The Tycho-Gaia Astrometric Survey (TGAS) has just been released
with approximately 2 million parallaxes, having typical reported
precisions of $\sigma(\pi)\sim 300\,\muas$.  Thus, while constituting only
a tiny fraction of the ultimate {\it Gaia} product, TGAS is by far the
largest and (with the exception of a tiny handful of {\it Hubble Space
Telescope} parallaxes, e.g., \citealt{benedict11}), 
the most accurate optical astrometric catalog now available \citep{gaia1,gaia2}.

In order to validate the TGAS catalog, it is natural to compare with
the best previously existing astrometric catalog, {\it Hipparcos}.
This is not ideal as TGAS has, overall, significantly better parallax
measurement compared to {\it Hipparcos}.  However, it is feasible, in
principle, because of the large number $(\sim 10^5)$ of overlapping entries
and because such validation requires only the measurement of two
error-renormalization parameters $(A,\sigma_0)$
\begin{equation}
\sigma_{\rm tgas}(\pi) = \sqrt{(A\sigma_{\rm int}(\pi))^2 + \sigma_0^2};
\quad (A,\sigma_0) = (1.4,0.20\,\mas)
\label{eqn:renorm}
\end{equation}
where $\sigma_{\rm int}$ and $\sigma_{\rm tgas}$ are respectively the
internal and reported (renormalized) errors, $\sigma_0$ is the
systematic error floor and $A$ is the renormalization factor.

However, it is notoriously difficult to calibrate superior data from
inferior data and, in particular, requires superb knowledge of the
error-structure of the inferior data set.

We therefore present an alternative method for calibrating the TGAS parallaxes,
which does not require any external astrometric data.

\section{{New Method to Calibrate TGAS}
\label{sec:method}}

As we have previously discussed, the RR Lyrae (RRL) period-luminosity
(PL) relation is expected to provide a powerful means to calibrate
astrometric data --- in particular {\it Gaia} data \citep{gk2016}.  In
that paper, we discussed measurement of the parallax zero point
$\pi_0$, which will not be feasible until the full {\it Gaia}
release.  However, as we show here, the RRL PL relation can {\it also}
be used as a tool to determine the TGAS error renormalization.  The
RRL PL relation (in, for example, $K$ band) has the form\footnote{In
principle, this relation also depends on metallicity.  Because
metallicity does not play a role in the current exercise as we show
below, we ignore it.  In principle, however, for a larger number of
objects, we expect that metallicity information can further improve
the errors reported here.}
\begin{equation}
M_K = M_{K,0} + B\log(P/P_0)
\label{eqn:pl}
\end{equation}
where $M_{K,0}$ and $B$ are parameters, and $P_0$ is an arbitrarily
chosen reference point.  In this work we choose $\log (P_0/{\rm day})
= -0.29$ so that $(M_{K,0},B)$ are roughly uncorrelated in the TGAS
data set.  Thus, if the period $P$, the mean magnitude $K$, and the
extinction $A_K$, are measured, the parallax can be predicted for a
given assumed $(M_{K,0},B)$
\begin{equation}
\pi_{\rm pred} = 10^{[K - A_K - (M_{K,0} + B\log(P/P_0))]/5 + 2}\,\mas
\label{eqn:pipred}
\end{equation}

Given an ensemble of RRL parallax measurements and 
{\it well understood} errors $(\pi,\sigma)_i$, one can
derive $(M_{K,0},B)$ by minimizing $\chi^2$,
\begin{equation}
\chi^2(M_{K,0},B) = \sum_i {[\pi_i - \pi_{{\rm pred},i}(M_{K,0},B)]^2\over \sigma_i^2}.
\label{eqn:chi2}
\end{equation}
In the present case, however, we will extract $K$ measurements from
the 2MASS catalog, which are at a single epoch.  Because RRL have
full amplitudes of about 0.5 mag in $K$ band, this introduces an
rms flux error of roughly 16\% into the relation, which corresponds
to a fractional error in the $\pi_{\rm pred}$ of 8\%.  Therefore, for
our particular case, Equation~(\ref{eqn:chi2}) must be rewritten
\begin{equation}
\chi^2(M_{K,0},B) = \sum_i {[\pi_i - \pi_{{\rm pred},i}(M_{K,0},B)]^2\over 
\sigma_i^2 + (\epsilon \pi_{{\rm pred},i})^2}
\qquad \epsilon=0.08 .
\label{eqn:chi2b}
\end{equation}
In principle one could augment $\epsilon$ (in quadrature) by the
photometric error, but this is negligible.  Likewise, if the
scatter in the relation were a major factor, then $\epsilon$
could be augmented by this as well.  We believe this is small 
\citep{madore13, beaton16}, and
in any case it is likely smaller than the uncertainly in our estimate
of $\epsilon$ itself, so we ignore it.

If we now consider that the nominal errors $\sigma_{\rm int}$ must be
modified as per Equation~(\ref{eqn:renorm}), then 
Equation~(\ref{eqn:chi2b}) must be modified as well:
\begin{equation}
\chi^2(M_{K,0},B,A,\sigma_0) = \sum_i {[\pi_i - \pi_{{\rm pred},i}(M_{K,0},B)]^2\over 
(A\sigma_{\rm int})^2 + \sigma_0^2 + (\epsilon \pi_{{\rm pred},i})^2}.
\label{eqn:chi2c}
\end{equation}

Using this equation, one can simultaneously determine the four
parameters $(M_{K,0},B,A,\sigma_0)$.  For each set of trial
error-renormalization parameters $(A,\sigma_0)$, one minimizes
$\chi^2$ over $(M_{K,0},B)$.  Then, following the method of
\citet{yee12,yee13}, we consider solutions to be acceptable if the cumulative
$\chi^2$ distribution (ordered by nominal errors) forms a straight
line with a slope of unity.  We note the following facts.  If the real
errors are subject to a systematic floor $\sigma_0$ but this is not
reflected in the formal errors, the cumulative distribution will
initially rise more quickly than the ``unity line'', before falling
back to the line (if parameters have been chosen so that $\chi^2/{\rm
  dof}=1$).  By contrast, if $\sigma_0$ is overestimated, then the
cumulative $\chi^2$ distribution will initially climb more slowly than
the line.  Thus the morphology of this comparison gives us critical
information on the error floor within the data.

\section{{Data}
\label{sec:data}}

We begin with the sample of 125 {\it Hipparcos}
RRab RRL used in the PG98 statistical
parallax study \citep{popow98a,popow98b,gould98}.  We find that all of
these stars are in 2MASS but only 115 are in TGAS.
We individually search for these stars in SIMBAD by {\it Hipparcos} number,
which allows us to determine their traditional names, e.g. ``RX Cet''.
We then search in \citet{fernley98}, which provides periods $P$,
reddening $E(B-V)$, and also metallicity [Fe/H].  We check to see
whether the fits are improved by including [Fe/H], but since they
are not, we ignore it.  Only nine of the 115 stars are missing from
this catalog.  However, two of these (Hip 42115 and 94134) do not
appear to be RRL, so we exclude them.  And one other (BX Dra) is
an eclipsing variable (so definitely having additional flux in the
2MASS aperture) and hence we ignore that as well.  We search GOOGLE
for periods of the remaining stars, succeeding for XZ Cet, SZ Hya,
BB Vir, AR Ser, V494 Sco, and failing for IN Ara.  This leaves
us with a sample of 111 stars.  We adopt $A_K = 0.3E(B-V)$.  For
the great majority of stars, this number is completely negligible
and for almost all the others it is quite small.  For the five
stars for which we lack this parameter, we set $A_K=0$. 

\begin{figure}
\centering
\includegraphics[width=95mm]{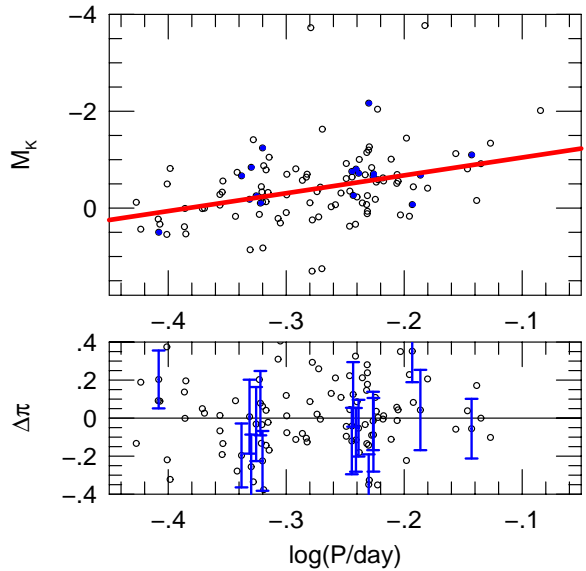}
\caption{Upper panel: RRL PL $K$-band relation derived from 108
RRab stars: $(M_{K,0},B)= (-0.35,-3.71)$.  Points are absolute
magnitudes derived from TGAS parallaxes, 2MASS photometry, and
(usually very small) $A_K$ based on \citet{fernley98} $E(B-V)$.
Method for deriving parameters is described in Section~\ref{sec:method}.
Lower panel: residuals of fit in parallax $\pi$.  Error bars are shown
for the 17 smallest-error parallaxes because these are the ones that
most constrain the systematics floor $\sigma_0=0.12$.  
See Figure~\ref{fig:cum}.  These 17 stars are marked in blue in the
upper panel.
\label{fig:pl}}
\end{figure}

We remove the three largest outliers to the fit (BB Vir, RV Cap, UY
Boo) for two reasons.  First, they are all $>2.9\,\sigma$ outliers, and
we do not expect such a high number arising from Gaussian noise in a
sample of $\sim 100$ objects.\footnote{With $\chi^2=(10.5,10.5,8.2)$,
these objects have probabilities of, respectively, about
$(0.1,0.1,0.6)$ of appearing in the Gaussian tails of 111 objects.
Thus, excluding the first two is clearly indicated, while the third
is basically indicated, but borderline.  However, whether the third
is included or excluded has almost no effect.  See main text.}
Second, one expects that at least a few stars will be ``corrupted'' by
real physical effects unrelated to the parallax measurement.
Retaining these objects obviously contaminates our measurement.  For
example, as noted above, BX Dra is known to have additional light in
the aperture from an eclipsing companion, so we know it should not be
included in our analysis.  Other stars undoubtedly have at least some
such light due to non-eclipsing companions.  Since RRL are giants,
this usually will not matter much but it may in a few cases.  Thus,
removing these objects is important not only for making the
the most accurate estimate of $(M_{k,0},B)$ (which is not the main
objective here), but also for making the most accurate estimate 
of $(A,\sigma_0)$ (which is).

\section{{Results}
\label{sec:results}}

\begin{figure}
\centering
\includegraphics[width=95mm]{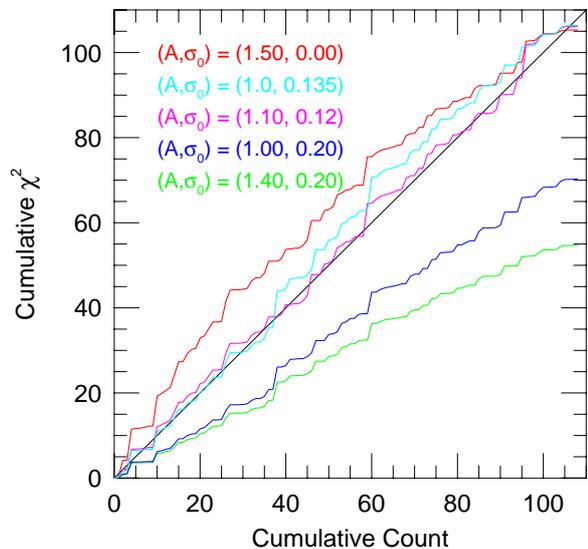}
\caption{Cumulative distribution of $\chi^2$ ordered from
smallest to largest formal error.  According to prescription of
\citet{yee12,yee13}, the parameters $(A,\sigma_0)$ entering 
Equation~(\ref{eqn:renorm}) should be adjusted so that this
distribution is a straight line with unit slope (black).
If there is no systematic floor (red), then cumulative $\chi^2$
rises too quickly at small errors (left).  If the rescaling factor
$A$ is set to its minimum physical value $A=1$ (cyan), the behavior
is acceptable, but not optimal.  The magenta curve 
$(A,\sigma_0)=(1.1,0.12\,\mas)$ is optimal.  The values adopted by
TGAS $(A,\sigma_0)=(1.4,0.2\,\mas)$ strongly overestimate the errors
(green), and this remains true even if one sets the rescaling factor
at its minimum physical value $A=1$, but leaves $\sigma_0=0.20$ (blue).
\label{fig:cum}}
\end{figure}

Figure~\ref{fig:pl} shows the RRL $K$-band PL relation and
Figure~\ref{fig:cum} shows cumulative distributions for various
choices of $(A,\sigma_0)$.  The red curve shows the result of assuming
$\sigma_0=0$.  The cumulative distribution definitely rises too
quickly.  The cyan curve shows the best case if we assume that only a
systematics term is needed (no rescaling factor: $A=1$).  In this case
$\sigma_0=0.135\,\mas$.  This would be acceptable.  However,
$(A,\sigma_0)=(1.1,0.12\,\mas)$ is clearly better.  The TGAS adopted
choice clearly falls below the ``unity line''.  Even if we accept the
TGAS $\sigma_0=0.2\,\mas$ and choose the minimum physically reasonable
value of $A=1$, the cumulative distribution still falls below the
line.  For completeness, if we eliminate only the 0 or 2 largest
outliers (as opposed to the three largest, which is our preference) we
obtain $A=1.35$ and $A=1.15$, respectively.  We caution, however, that
such procedures (particularly the former) would likely only be injecting
corrupted measurements into the final result rather than respecting
pure statistical protocol.

\section{{Discussion}
\label{sec:discuss}}

We provide an independent method to calibrate the TGAS catalog errors
that does not rely in any way on previous generation astrometric data.
This purely photometric method of the standard candles avoids the issues
associated with relying on lower quality data and
ultimately yields more precise values for the TGAS catalog.

The method presented here is likely to be more accurate than one
based on comparison to {\it Hipparcos} parallaxes.  This is particularly
true for measuring the zero-point floor $\sigma_0$, for which we find
$\sigma_0=0.12\,\mas$ and the TGAS team found $\sigma_0=0.20\,\mas$.

The {\it Hipparcos} catalog very likely
has systematic errors at the $\sim 0.1\,\mas$ level.
This is
well below the typical statistical error for individual stars, 
so the only place that these systematics have surfaced is in the
measurement of the distance to the Pleiades, which was based on
combining measurements of many stars.  The original {\it Hipparcos}
parallax was larger than the ``traditional'' value by almost 1 mas,
and it was suggested at that time that this could be due to correlated
errors \citep{pin98,sod98}.  \citet{ng99} demonstrated strong evidence
for correlated {\it Hipparcos} errors in the Hyades field which, by
chance, they showed did not lead to an error in the distance estimate.
However, after reanalyzing the {\it Hipparcos} data, \citet{van09}
could find no internal evidence of these correlations and published
a similar Pleiades distance measurement as originally with yet smaller
error bars.  This conflict was resolved by TGAS in favor of the
``traditional'', longer Pleiades distance \citep{gaia1}.

While the distance to the Pleiades is settled, the cause of the
{\it Hipparcos} error in its estimate of this distance is not.
It may be that the problem is entirely explained by correlations,
but these may also be masking other problems.  In particular, since
the cause of these correlations (if they are in fact the root cause)
have not been tracked down, it cannot be assumed that the {\it Hipparcos}
error profile is understood at a level well below the precision of
its measurement.

By contrast, the external inputs into RRL PL relation ($P$, $K$, $A_K$)
are quite well understood.  Therefore, the approach of the current paper
appears more secure and in any case independent from the purely astrometric approach.

One important difference between the sample studied here (108 RRL)
and the one studied by the {\it Gaia} team ($10^5$ {\it Hipparcos} stars)
is that the RRL have intrinsically similar colors while the {\it Hipparcos}
stars cover a full range of stellar colors.  In principle, this
difference could be important since the {\it Gaia} team did not
attempt to correct TGAS for color-dependent astrometric deviations (as they
will for subsequent releases).

This issue clearly deserves further investigation, a path to which
we outline below.  However, if there are color-dependent systematic
errors, these are likely to be larger for RRL than average stars
because RRL have a larger color offset relative to the mean
reference frame set by other stars.

Nevertheless, this question can be further investigated
by applying the same method that we have used here, but to Cepheids.
Cepheids lie in the same instability strip and so, like RRL,
they are systemically bluer than other stars.  However, unlike RRL
they virtually all lie in the Galactic plane.  Moreover, they
are more luminous than RRL and so are typically seen
at greater distances and so through more dust.  In our sample
of 108 RRL, there are only 19 stars with $E(B-V)>0.1$ and
only four of these have $E(B-V)>0.2$.  This is not enough
to probe a broad range of observed colors.  By contrast, Cepheids
will probe a broad range.  Note in particular that while 
distant Cepheids provide relatively little information about the
PL relation, they can provide excellent information on
TGAS error characterization.  This is because, once the
PL relation is determined from nearby stars, the parallaxes
of distant stars can be determined photometrically to much
higher precision than the parallax errors.  This is the same
principle used in the \citet{gk2016} method of measuring
$\pi_0$.


\acknowledgments
This work was supported by NSF grant AST-1516842.


\end{document}